\documentclass[utf8]{frontiersSCNS} 
\usepackage[onehalfspacing]{setspace}

\def\keyFont{\fontsize{8}{11}\helveticabold }
\def\firstAuthorLast{Czerny {et~al.}} 
\def\Authors{Bozena Czerny\,$^{1,*}$, Yan-Rong Li\,$^{2}$, Justyna Sredzinska\,$^{3}$, Krzysztof Hryniewicz\,$^{3}$, Swayam Panda\,$^{1,3}$, Conor Wildy\,$^{1}$, and Vladimir Karas\,$^{4}$}
\def\Address{$^{1}$Center for Theoretical Physics, Polish Academy of Sciences, Al. Lotnikow 32/46, 02-668 Warsaw, Poland \\
$^{2}$Key Laboratory for Particle Astrophysics, Institute of High Energy Physics, Chinese Academy of Sciences, 19B Yuquan
Road, Beijing 100049, China\\
$^{3}$Copernicus Astronomical Center, Polish Academy of Sciences, Bartycka 18, 00-716 Warsaw, Poland\\
$^{4}$Astronomical Institute, Czech Academy of Sciences, Bocni II 1401, CZ-141 00 Prague, Czech Republic}
\def\corrAuthor{Bozena Czerny}

\def\corrEmail{bcz@cft.edu.pl}

\begin{document}
\onecolumn
\firstpage{1}

\title[Self-consistent dynamical model of the BLR]{Self-consistent dynamical model of the Broad Line Region} 

\author[\firstAuthorLast ]{\Authors} 
\address{} 
\correspondance{} 

\extraAuth{}

\maketitle

\begin{abstract}
We develop a self-consistent description of the Broad Line Region based on the concept of a failed wind 
powered by radiation pressure acting on a dusty accretion disk atmosphere in Keplerian motion. 
The material raised high above the disk is
illuminated, dust evaporates, and the matter falls back towards the disk. This material is the source of emission lines. 
The model predicts the inner and outer
radius of the region, the cloud dynamics under the dust radiation pressure and, subsequently, the gravitational field
of the central black hole, which results in asymmetry between the rise and fall. Knowledge of the dynamics allows us to predict the shapes of the emission lines as functions of the basic parameters of
an active nucleus: black hole mass, accretion rate, black hole spin (or accretion efficiency) and the viewing angle with respect to the symmetry axis. Here we show preliminary results based on analytical approximations to the cloud motion.
\tiny
 \keyFont{ \section{Keywords:} emission lines, active galactic nuclei, Broad Line Region, accretion disk, black hole} 
\end{abstract}

\section{Introduction}

The Broad Emission Line Region (BLR) is the key ingredient of most active nuclei, and its true nature remains illusive. Huge observational progress allowed the accumulation of a lot of 
constraints/requirements for the material which is the source of emission lines. This (clumpy ?)
wind emission is mostly powered by the emission from the most central parts, with (possibly) some 
contribution from mechanical heating. The distribution of the emitting material is rather flat since BLR
clouds are rarely seen in absorption despite covering a relatively large fraction of the sky as seen from the nucleus,
which is required by the total line luminosity. The study of emission line variability allows measurement of the distance to the BLR, and shows that in general the region is extended, with High Ionization Lines (HIL) forming closer in, and Low Ionization Lines (LIL) forming further down \cite{souffrin1986,gaskell2009}. The motion is predominantly
Keplerian, which opened a way to use the BLR for black hole mass measurements \cite{wandel1999}. The wavelength-resolved 
reverberation mapping now allows for a few well-studied sources to provide an insight into the velocity
field, confirming predominantly Keplerian motion but with traces of inflow as well as outflow \cite{grier2013}.

The outflow suggests that the emitting material is connected with the cold accretion disk present in the nucleus, 
and a disk wind. There are also observational arguments for the co-existence of the cold disk and BLR.

Several mechanisms are known to drive winds (radiation pressure driven winds; e.g. \cite{elvis2012}; magnetically 
driven winds, e.g. \cite{fukumura2015}; pressure driven winds, e.g. \cite{fukue2004}, and the references therein),
 and a combination of all effects is likely to contribute although
various mechanisms could dominate at various distances from the black hole, as the local densities and ionizing flux
changes with the disk radius. Inflow might seem less expected, but the winds in some parameter space are actually failed 
winds, not escaping to infinity, and reversing to inflow. Also the thermal instability in the circumnuclear material
may lead to selective inflow of the denser phase e.g. \citep[e.g.]{elvis2017}.

This tremendously complicated region, however, produces a strikingly good correlation between
the BLR size and the source monochromatic luminosity. More specifically, this is the relation between
the delay of the H$beta$ line with respect to the 5100 \AA~ continuum, which in addition is not the continuum driving this line.
This relation is now observationally studied in a broad luminosity range as well as in redshift \cite{peterson2004,bentz2013,shen2016}.

Failed Radiatively Accelerated Dusty Outflow (FRADO) model \citep{czhr2011,czerny2015} offers a natural way to understand this relation. Other BLR models, like the one developed by \cite{netzer2010} are parametric, they do not uniquely predict the radial extension of the BLR without referring to an observational scaling while FRADO gives the location of the LIL part, at the basis of the dust microscopic properties. Below
we briefly outline the model itself, the dynamics of the clouds and the exemplary line profiles. The model parameters
are just the global parameters of the stationary accretion flow in the nucleus: the black hole mass, the accretion rate,
the spin (or accretion efficiency) and the viewing angle towards the nucleus. 

\section{FRADO model}

The model is based on the known observational fact that stellar winds are much more prominent in cooler stars with dusty
atmospheres than in the hotter stars. Since the accretion disk in a given source has a broad temperature range, there exists
a distance from the black hole where the effective temperature of the disk drops below the dust sublimation temperature. There the
material is risen efficiently above the disk by the local radiation pressure. However, with increasing disk height the dusty 
material is exposed to the irradiation by the central regions, the plasma temperature rises above the sublimation temperature, 
the dust evaporates, and the radiation pressure vanishes, so the clouds move by following ballistic motion, first up and then
down towards the disk surface. We neglect here the other sources of the radiation pressure, like line-driven outflow.  In the 
present work we aim at catching the most characteristic properties of the model in a broad parameter range, so we introduce some
simplifications which allow us to describe the cloud dynamics analytically. The basic model parameters are simply the 
global parameters of an active nucleus, i.e. the black hole mass, accretion rate, black hole spin (or accretion efficiency) and the viewing angle with respect to the symmetry axis. Other parameters, like the dust sublimation temperature and the dust opacity should result, at least in principle, from the basic physics. The scenario is outlined in Fig. 1.

\section{cloud dynamics}

We assume that the dust opacity can be described in grey approximation as a wavelength-independent value, and we neglect its dependence on the temperature as well, as long as the temperature is lower than the dust sublimation temperature. In the presence of the dust within the cloud, the cloud motion can be approximated as
\begin{equation}
{dv \over dt} = - {G M_{BH} \zeta \over r^3} + {G M_{BH} \over r^3}H_{disk}\bigl({\kappa_P \over \kappa_R} - 1\bigr).
\end{equation}
where $\zeta = z - H_{disk}$ is the distance measured from the disk surface. Here $H_{disk}$ is the disk thickness, $M_{BH}$ is the black hole mass, and $r$ is the current disk radius. If the disk is dominated by the radiation pressure, $H_{disk}$ is constant and its value is given by accretion rate $\dot M$ as $ H_{disk} = {3 \kappa_R \dot M \over 8 \pi c}$. The clouds also perform circular motion with the local Keplerian velocity. If both the Planck mean $\kappa_P$ and the Rosseland mean $\kappa_R$ are 
constant and the dust does not evaporate, this equation can be easily integrated analytically to get
\begin{equation}
\label{eq:z_normal}
\zeta = H_{disk} ({\kappa_P \over \kappa_R} - 1) \bigl[ 1 -\cos \Omega_K t \bigr],
\end{equation}
where $\Omega_K$ is the local Keplerian angular velocity. The cloud does perform an oscillatory motion from $\zeta = 0$ to $\zeta_{max} = 2 H_{disk} (\kappa_P/\kappa_R - 1)$. The maximum cloud velocity in the vertical direction is a constant fraction of the local Keplerian velocity so it depends on the radius. The cloud maximum height in this approximation is independent of the radius, so the nearest clouds intercept most of the central radiation. Both the velocity and the height scale with the accretion rate, $\dot M$, in physical units, i.e. scales both with the black hole mass and accretion rate in Eddington units. Such a description is only valid when the dust within the cloud survives. However, close to the inner radius of the BLR evaporation is important, and the condition for dust evaporation can be formulated as
\begin{equation}
\zeta_{evap} = {\sigma_B T^4_{dust} 4 \pi r^3 \over \eta \dot M c^2} - {3 \over 2}{1 \over \eta} r_g,
\end{equation}
where $r_g = GM/c^2$ is the gravitational radius and $\eta$ is the accretion efficiency which connects the accretion rate to the
bolometric luminosity. If $\zeta > \zeta_{evap}$ the pressure term in the equation of motion disappears. The further motion of the cloud up and down can still be calculated analytically but the formulae become more complicated due to asymmetry between part of the rise supported by the dust and the second part of the rise and subsequent fall of the cloud. The schematic structure of the BLR parts is shown in Fig. 2, left panel, and the asymmetry in motion is illustrated in Fig. 2, right panel.  

The inner radius of the BLR is set by the condition of $\zeta_{max} = 0$, and the outer radius of the BLR is given by $\zeta_{evap} = r$, i.e. by the condition that the dust survives the irradiation even high above the disk mid-plane. This region in general consists of two parts: the inner part where both dusty and dustless clouds are present and the outer region where $\zeta_{max}$ is smaller than $\zeta_{evap}$ and only dusty clouds exist. The transition between the A/B and C zones is set by the condition $\zeta_{max} = \zeta_{evap}$. This structure is shown schematically in Fig. 2. The extension of the zones depend on the mass, accretion rate and the accretion efficiency.  

The position of the transition radius, $r_{AC}$, can be used to evaluate the basic expected trends of the model. The ratio of $r_{AC}$ to the inner radius increases with $\dot M^{1/3}/M^{1/3}$, i.e. the dimensionless accretion rate, thus the emission comes from a broader range of radii and the double-peak character of the line profile is less expected with the rise of the Eddington ratio.  


\section{line profiles}

We now assume that the line is emitted as a monochromatic line, later broadened by the vertical and rotational motion. The line emissivity is assumed to be directly proportional to the incident flux, i.e. the efficiency of radiation reprocessing characteristic for a specific emission line is neglected in the current model. Thus the local emissivity at a given radius is simply assumed to be proportional to the local disk surface element, $dr d\phi/r^2$, where $dr$ and $d\phi$ are the elements of the radial and azimuthal grid. The cloud distribution is sampled in all three dimensions, the distribution in the $z$ direction is consistent with the cloud motion, and the Moon-type effect in the cloud emission is taken into account. The computations of the line profile neglect the Doppler boosting and gravitational reddening but the profiles are computed for an arbitrary viewing angle of an observer. The final line shape is then calculated numerically. 

In the final computations we adopted at present the following values of the constants: $T_{dust} = 1500$ K, $\kappa_P = 8.0$ cm$^{-2}$g$^{-1}$,  $\kappa_R = 4.0$ cm$^{-2}$g$^{-1}$, and the accretion efficiency $\eta = 0.1$, corresponding to a moderate black hole spin.  

Fig. 3 (left) shows the dependence of the predicted line profiles on the accretion rate, for a fixed black hole mass $M = 10^8 M_{\odot}$ and accretion efficiency of 0.1 but for three values of the accretion rate corresponding to the Eddington ratio of 0.01, 0.1 and 1, for a viewing angle of 30$^{\circ}$. The line always shows a two peak structure but this disk-like component is much stronger in the case of a low Eddington ratio than in the case of a high Eddington ratio. This change reflects two trends: (i) with the rise of the accretion rate the BLR moves outwards so the line in general becomes narrower (ii) the ratio of the maximum vertical to the local azimuthal velocity also rises with the accretion rate so that the contribution of the vertical motion to the line profile is more important. This trend well reproduces the fact that the lines in Narrow Line Seyfert galaxies do not show the double peak structure, while this double peak structure is clearly visible in low Eddington ratio sources, particularly in the variable part of the spectrum. In Fig. 3 (right) we show a sequence of the solutions for a different black hole mass and the Eddington ratio 1.0. We see that the line shape still shows some double-peak structure, but it is much smaller in high Eddington ratio sources than in the low Eddington ratio sources. The trend depends also on the black hole mass itself (see Fig. 3, right panel), with larger black hole mass showing more double-peak behaviour. It is consistent with the division of the Seyfert 1 galaxies into two classes at FWHM of 2000 km s$^{-1}$ (Seyfert 1 and Narrow Line Seyfert 1 objects) and similar division into class A and B in quasars at much higher FWHM of 4000 km s$^{-1}$ (type A and type B quasars). 

\section{Discussion}

The results above show the first very simplified attempt to use the FRADO model for explaining the structure of the BLR. It has a number of properties consistent with the observed trends, like narrower lines for higher Eddington rate sources and smaller black hole mass. However, the model is not yet ready for a detailed comparison with the AGN spectra, as the Baldwin effect is not expected from the current version of the model. First of all, the exact values of the physical constants describing the dust properites should be included in the model. The mean time delay suggesting dust values $\sim 900 $K is most likely incorrect since the measured delay does not simply translate to the BLR inner radius. Spectroscopic studies indicate larger hot dust temperatures, which will give broader line profiles. At a later step cloud shielding should be included, combined with a better description of the radial weight of the line contribution due to geometrical setup. The disk thickness should include the transition from the radiation-pressure supported inner part to the gas-supported outer part which makes the disk shape more complex than the constant disk thickness used in the current work. Finally, the real cloud emissivity should be included, and the radiation pressure should be described much more carefully, taking into account the wavelength-dependence of the radiation pressure acting on dust, dust composition and the radiation pressure due to the absorption in lines. Those last aspects are particularly complicated, although some progress have been already done in this direction (\cite{waters2017,krolik2016,gallagher2015,hoenig2017}.   

The picture outlined in this paper is basically stationary. In order to include the time-dependent behaviour we would have to introduce the time-dependent emission from the inner disk. Thermal and/or magnetic field fluctuations are required to model the usual red noise variability of an AGN continuum, and they would lead to delayed line responce but most likely without strong change in the line shape. Observed longer systematic trends in the line shapes (e.g. \cite{li2016,bon2016,sredzinska2017}) require additional factors perturbing the disk symmetry like a secondary black hole, accretion disk precession or spiral waves present in the disk due to self-gravity effects.   

\section*{Conflict of Interest Statement}

The authors declare that the research was conducted in the absence of any commercial or financial relationships that could be construed as a potential conflict of interest.

\section*{Author Contributions}

BC was responsible for the idea that resulted in the paper and most of the text, KH contributed to the concept and the text, JS derived the basic formulae for dusty clouds, later modified by VK, SP and CW for the dust evaporation effect, YR contributed to the line profile computations. 

\section*{Funding}
Part of this work was supported by Polish grant Nr. 2015/17/B/ST9/03436/.

\section*{Acknowledgments}
We are grateful to Jian-Min Wang for many helpful discussions of the project during the stay of BC in Beijing.

\bibliographystyle{frontiersinSCNS_ENG_HUMS} 
\bibliography{czerny}


\section*{Figure captions}

\begin{figure}[h!]
\begin{center}
\includegraphics[width=10cm]{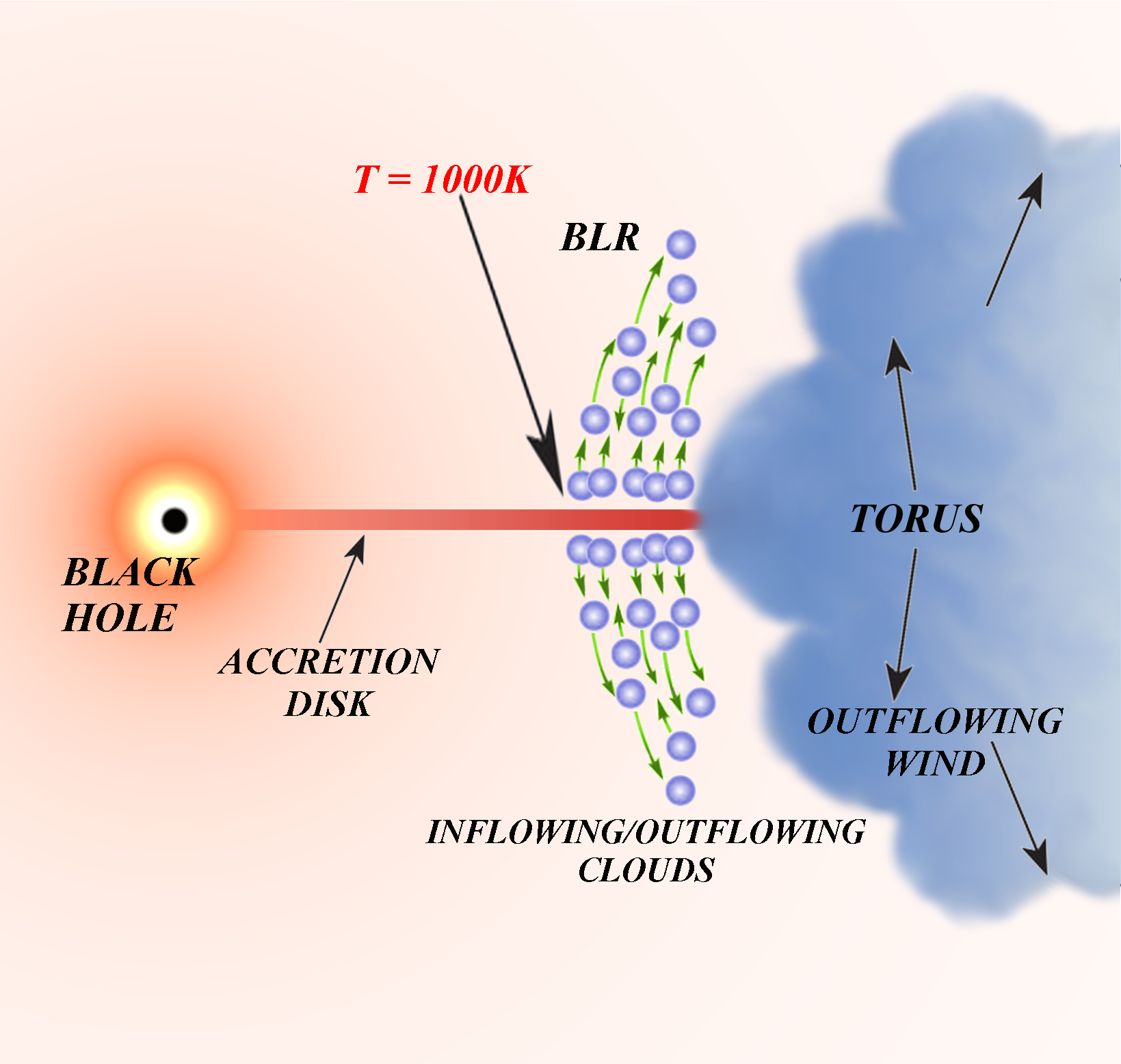}
\end{center}
\caption{Concept of the FRADO model for the BLR.}\label{fig:1}
\end{figure}

\begin{figure}[h!]
\begin{minipage}[b]{.5\linewidth}
\centering\includegraphics[width=10cm]{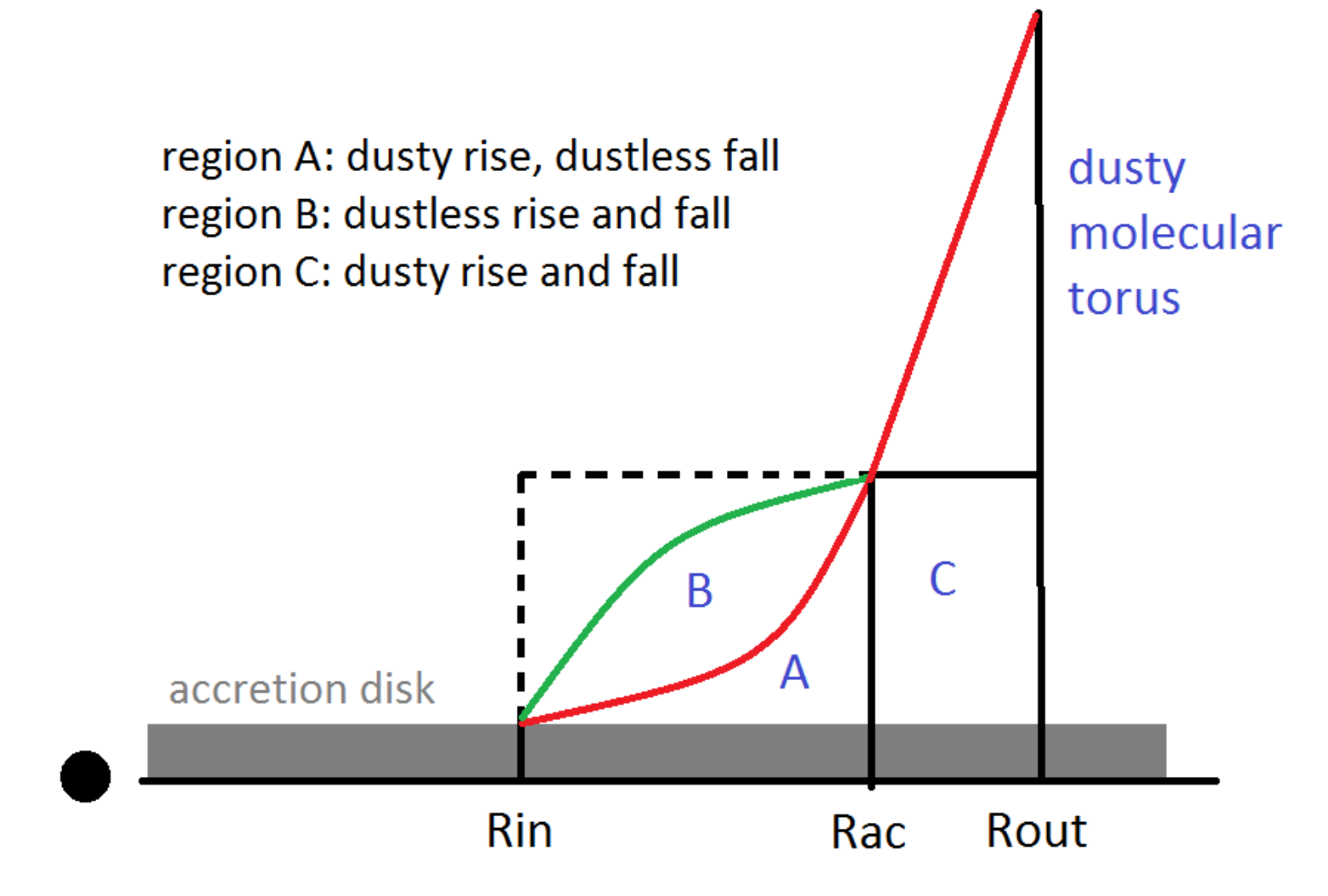}
\end{minipage}%
\begin{minipage}[b]{.5\linewidth}
\centering\includegraphics[width=7cm]{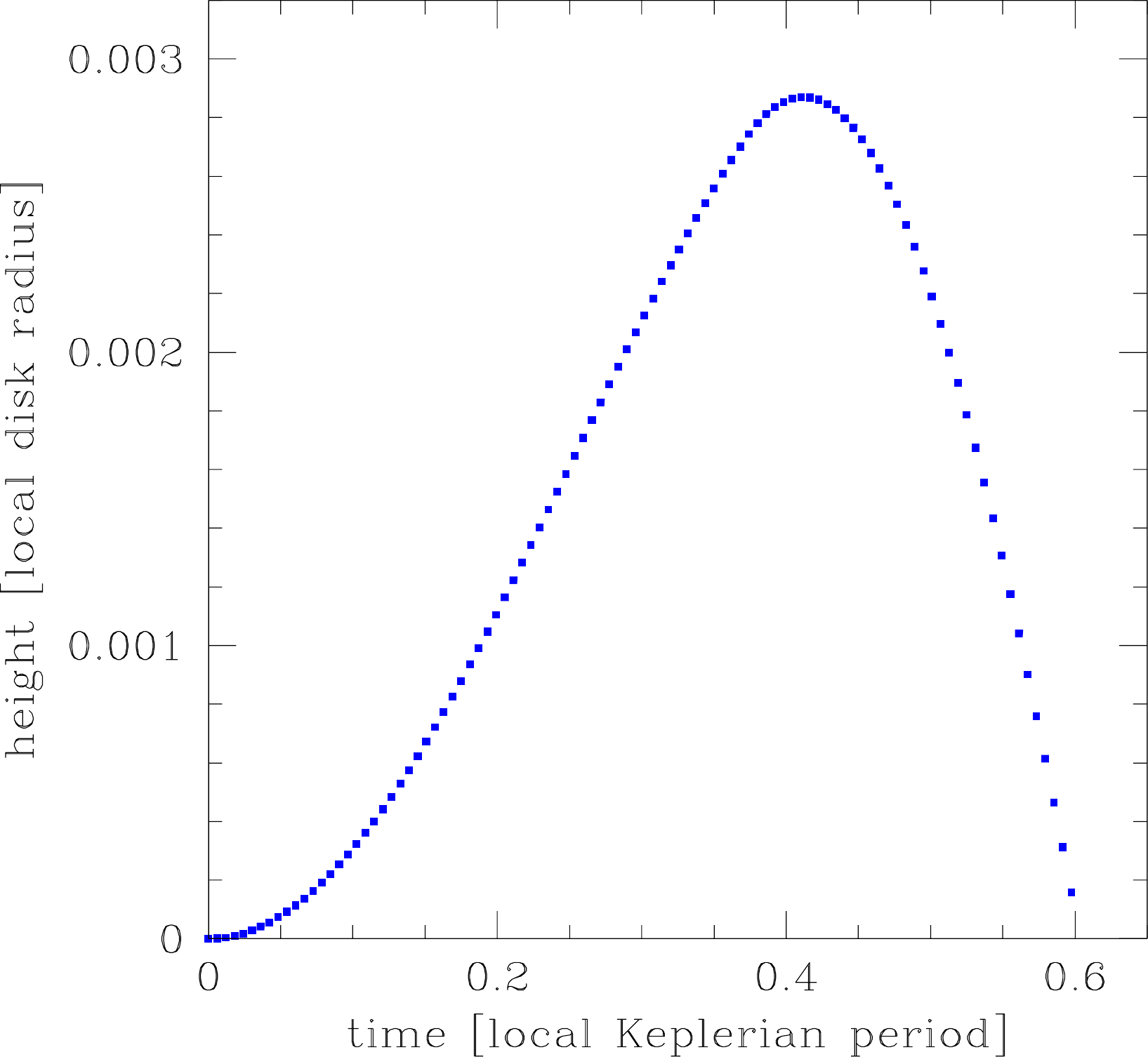}
\end{minipage}
\caption{ Schematic drawing of the BLR region (left panel) and an example of a single cloud motion in the A/B region showing clear asymmetry in the rise and fall motion due to the dust evaporation (right panel).}\label{fig:2}
\end{figure}

\begin{figure}[h!]
\begin{center}
\includegraphics[width=15cm]{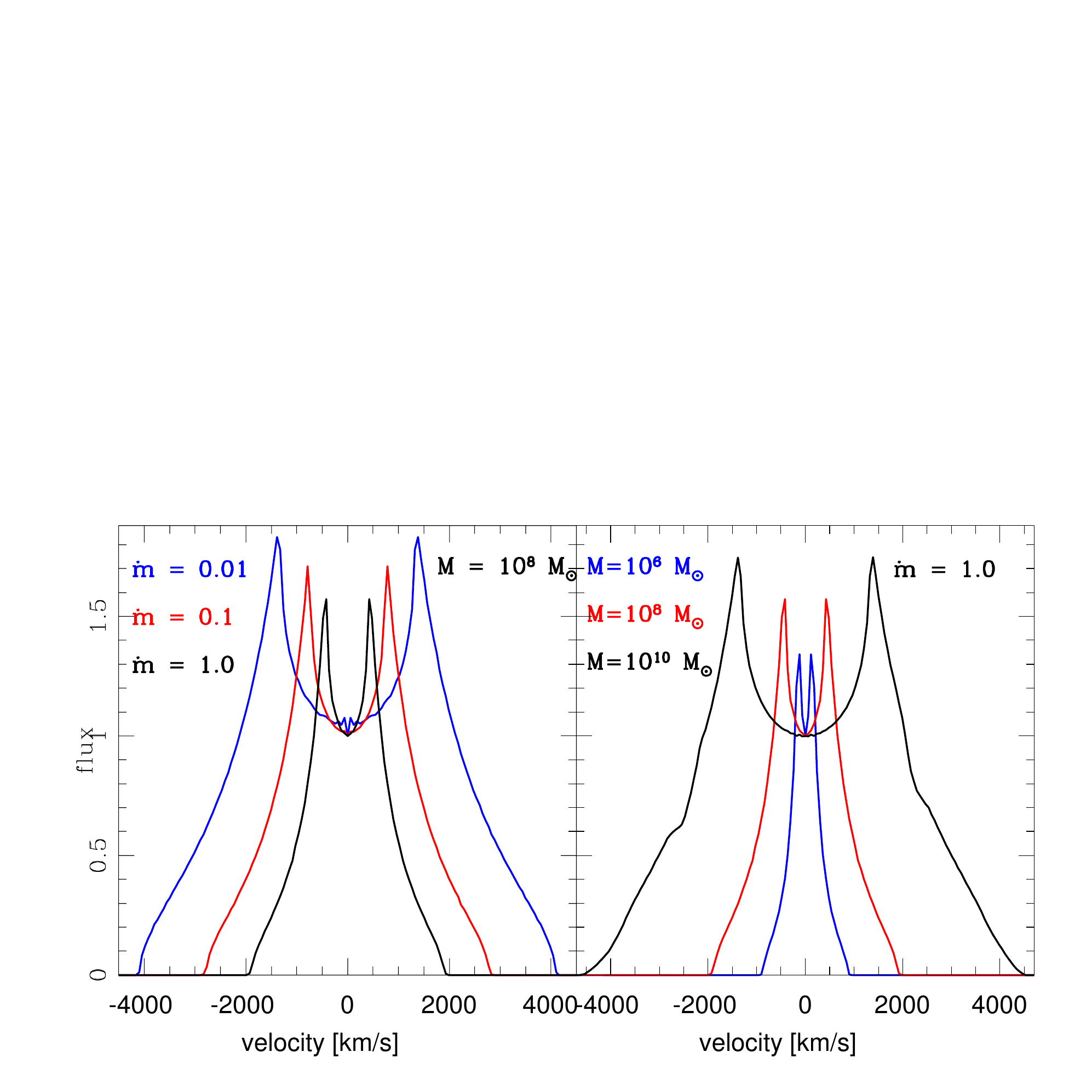}
\end{center}
\caption{An example of the dependence of the line profile on the accretion rate, for a black hole mass of $10^8 M_{\odot}$ (left panel), and on the black hole mass for the Eddington accretion rate (right panel). The viewing angle is fixed at 30 deg. The inner and outer BLR radii calculated for the model are: $1.98 \times 10^{16}$ cm and $ 1.87 \times 10^{17}$ cm, $4.26 \times 10^{16}$ cm and $ 5.91 \times 10^{17}$ cm, $9.18 \times 10^{16}$ cm and $ 1.87 \times 10^{18}$ cm in the left panel, and $4.26 \times 10^{15}$ cm and $ 1.87 \times 10^{17}$ cm, $9.18 \times 10^{16}$ cm and $ 1.87 \times 10^{18}$ cm, $1.98 \times 10^{18}$ cm and $ 1.87 \times 10^{19}$ cm in the right panel. Line flux is normalized to 1 at zero velocity.}\label{fig:3}
\end{figure}

\end{document}